[1994/12/01]
\documentclass[draft]{article}
\title{Critical Nucleation in Colossal Magnetoresistance }

\author{   A.  Barra\~n\'on  
\footnote{ Universidad Aut\'onoma Metropolitana. Unidad Azcapotzalco.
Av. San Pablo 124, Col. Reynosa-Tamaulipas, Mexico City. email: bca@correo.azc.uam.mx } ;
 J. A.  L\'opez
\footnote{Dept. of Physics, The University of Texas at El Paso. El Paso, TX, 79968  }    ;
C. Dorso
\footnote{ Dep. de F\'{\i}sica, Universidad de Buenos Aires. Buenos Aires, Argentina}    ;
 Fr.de L. Castillo 
\footnote{ Depto.de F\'{\i}sica, ESFM-IPN, Zacatenco, Mexico City}     }

\date{October, 20th, 2002}

\usepackage{amsmath,amsthm}

\chardef\bslash=`\\ 
\newcommand{\ntt}{\normalfont\ttfamily}

\theoremstyle{definition}

\theoremstyle{remark}

\newcommand{\eval}[2][\right]{\relax
  \ifx#1\right\relax \left.\fi#2#1\rvert}

\newcommand{\envert}[1]{\left\lvert#1\right\rvert}

\begin{document}
\maketitle
\markboth{Sample paper for the {\protect\ntt\lowercase{amsmath}} package}
{Sample paper for the {\protect\ntt\lowercase{amsmath}} package}
\renewcommand{\sectionmark}[1]{}

\abstract

Critical exponents have been obtained for a 3D spin particle system. 
Clusters are formed and system reaches a critical behavior 
when fragment size distribution follows a power law, as predicted by 
Fisher Liquid Droplet Model. Also, spontaneous magnetization critical
 temperature is in agreement with other theoretical studies. System 
evolution is reproduced via a genetic algorithm that performs minimal 
genetic fluctuations until a stationary state is attained. 
Critical exponents are in the range of those belonging 
to Heavy Ion collisions previously reported, and therefore belong to 
the same universality class.


\section{Introduction}

Theoretical and experimental studies have shown that homogeneous nucleation is
 fundamental for nuclear systems meanwhile inhomogeneous nucleation 
is relevant for macroscopic systems. According to Grassberger universality
 hypothesis, models exhibiting a continuous phase transition to an absorbing
 state belong to the same universality class as directed percolation \cite{Grass}. 
Rossi ${\it et. al.}$ have introduced a conservative gas lattice with 
a short range stochastic interaction that exhibits a continuous phase 
transition to an absorbing state for a critical value of the density of 
particles \cite{Rossi}. 

   In this study we obtain critical exponents of the Colossal Magnetoresistance (Fig. 1),
 namely $\tau = 2.38$ and $\beta= 0.38$ (Fig. 2), which are equal to those recently 
reported by Kudzia et. al for Heavy Ions collisions \cite{Kudzia}. Critical temperature is 
in the range of 4.5K (Fig. 3) and nuclear collisions critical temperature is in the range
 of 4.5 MeV. Power law critical exponent is the same as the one found by Mader
 ${\it et. al.}$ in his study about reducibility and thermal scaling in the Ising model, 
where a critical exponent $\tau = 2.39$ was obtained \cite{Mader}. 
Therefore both phenomena belong to the same universality class.

  The manuscript introduces the basic ingredients required to 
study critical phenomena thermodynamics in section 2. Critical exponents for 
colossal magnetoresistance are computed in section 3. Critical exponents for
 collision Molecular Dynamics  simulations are obtained in section 4. Finally,
 conclusions are established in section~\ref{CP}.
 
\section{Nucleation Critical Exponents}\label{NT}

 The  general shape of a Nuclear Model phase diagram is shown in fig. 4, 
where nucleus is considered a collection of protons and neutrons into a
 potential well, interacting via a phenomenological 
interaction~\cite{lopezdorso,kapusta}.

Stability conditions for asymmetric nuclear matter are given by:
   
\begin{equation}
\biggl( \frac{\partial E}{\partial T} \biggr)_{\rho,\delta} > 0,
\end{equation}

\begin{equation}
\biggl( \frac{\partial P}{\partial \rho} \biggr)_{T,\delta} \ge 0,
\end{equation}

\begin{equation}
\biggl( \frac{\partial \mu_n }{\partial \delta} \biggr)_{P,T} \ge 0,
\end{equation}
where $E$,$P$ and $\mu_n$ are the energy per nucleon, pressure and neutron
 chemical potential. These three equations are related to thermodynamic, 
mechanic and diffusive equilibriums.

According to fluctuations theory~\cite{landau}, 
the probability of having a liquid droplet of radius $r$ and $A$ 
vapor nucleons at temperature $T$, is given by
 $P_r(A)\propto e^{-\Delta G/T}$, with $\Delta G$ equal to the
 difference between the Gibbs Free energy for both phases. 

Using the $\Delta G$ given by the surface, curvature and bulk terms, 
we obtain~\cite{lopezdorso}:
\begin{eqnarray*}
P_{r}(A)=Y_{0}A^{-\tau }e^{-\left[ \left( \mu _{l}-\mu _{g}\right)
A+4\pi r^2_{0}\sigma (T)A^{2/3}\right] /T}  \ ,
\end{eqnarray*}
with $Y_0$ a normalization constant. The graph of this functional form in 
the supersaturated region has a U shape. Nevertheless, in the coexistence region,  
  $\left( \mu _l-\mu_g\right) =0$, and $P_r(A)=Y_0 A^{-\tau}\exp [{-4\pi
r_0^2\sigma(T)A^{2/3}/T}]$, 
describing a power law times an exponential decay relevant for 
 big masses. Finally, in the critical point 
$\left( \mu _l-\mu _g\right) =0$, 
since liquid and gas are indistinguishable at this point, surface tension is null,
$\sigma (T_c)=0$, and hence the following distribution is expected:
\begin{eqnarray}\label{powerl}
P_r(A)=Y_0A^{-\tau }  \ ,
\end{eqnarray}
which is a power law, and as such, scale free.

The exponent $\tau$ of the droplet size distribution~(\ref{powerl}), 
is known as the critical exponent since it is a dimensionless constant 
with a common value for different systems.
In this case $\tau$ can be obtained from a power law fit on the mass 
distribution, as will be shown hereby.

The Fisher liquid droplet model ($FDM$) for nucleation ~\cite{fisher} 
refines the probability~(\ref{powerl}) 
to obtain a critical liquid droplet mass distribution,
normalized to the size of the system:
\begin{equation}\label{nA}
n_{A} =q_{o}A^{-\tau}
\end{equation}
with a proportionality constant $q_{o}$ that can be obtained from the first moment,
 $M_{1}=\sum_{A}n_{A}A$, of the normalized mass distribution,
 ({\it i.e.} $M_{1}=1$). Since at the critical point $q_{0} =n_{A}A^{\tau}$,
the moment comes out to be $M_{1}=q_{o}\sum_{A}A^{(1-\tau)}=1$, and $q_{o}$ 
could be obtained through the following relation:
$q_{o}={1}/{\sum_{A}A^{(1-\tau)}}$. 
Notice that, since in the critical point any liquid droplet can be excluded, the sum
 is carried out on the clusters belonging to the gas phase.

Critical multiplicity provides a best $\chi^2$ fit ~\cite{dorlop2001}. This technique is used in
 section~\ref{cmr} with  data generated by the genetic algorithm describing colossal
 magnetoresistance and in section~\ref{mol} on Molecular Dynamics.

Critical exponents are given by the following expressions \cite{Gil1}:

\begin{equation}
M_2 \sim { \envert{ \epsilon}}^{- \gamma } 
\end{equation}

\begin{equation} A_{max} \sim { \envert{ \epsilon}}^{ \beta } 
\end{equation}
\begin{equation}  n_{A} \sim {A}^{ - \tau } \end{equation}

where  $m=m_c$. And satisfy the following equation:

\begin{equation}
\tau = 2 + \frac {\beta} {\beta+\gamma}
\end{equation}

  Asymptotic mass distribution moments are given by:
\begin{equation}    m_k^{(j)} = \frac { \sum_A A^k n^{(j)} (A) } {A_{tot}} 
\end{equation} 
when $k=2$ this moment is proportional to the isothermal compressibility.

   Assuming scaling in fragment size distribution, close to the critical point:
\begin{equation}  M_k \sim   \envert{ T - T_c  }^{ (-1- k + \tau )/ \sigma} ,
\end{equation}
 where: $ 2 < \tau < 3$.

\section{Colossal magnetoresistance critical exponent $\tau$ }\label{cmr}

   Flory introduced percolation in the context of polymer gelation, and it has
 been used ever since in a great variety of techniques \cite{Flory}. For a wide
 spectra of spin models there is a mapping of a equivalent graphical
 representation with a percolating transition related to the spin model phase
 transition \cite{Chayes}. 

In a percolation cluster the activated bonds extend from one side of the 
lattice to the opposite side. For infinite systems, there is a well defined 
``critical occupation probability'' $p_{c}$, above which the probability of
 finding a percolation cluster is $1$, meanwhile below $p_{c}$ 
this probability is $0$. For finite lattices, this transition is soft, {\it i.e.} 
the probability of finding a percolation cluster is different of $0$
for any occupation probability.

 Mader ${\it et. al.}$ have found thermal scaling and reducibility properties in the
 Ising model, obtaining a value of $\tau = 2.39$ \cite{Mader}. Chayes et.al. have
 shown that infinite clusters scale with a critical exponent $\tau -2=1/2$ and
 finite clusters scale with a critical exponent $\tau -2=1/3$ \cite{Chayes1}. 
In the percolation model the weight of a given configuration C with n bonds is given by:
\begin{equation}
W(C)=p^n(1-p)^{N-n}
\end{equation}
where N is the number of vertices in the lattice. Close to the percolation threshold,
 the critical behavior is characterized by the following critical exponents:
\begin{equation}
P_{\infty}=1- \sum s n(s,p) \sim \envert{p-p_C}^{\beta}
\end{equation}
and:
\begin{equation}
S(P)=\sum s^2 n(s,p) \sim \envert{p-p_C}^{\gamma}
\end{equation}

The cluster distribution satisfies the following scaling relation:
\begin{equation}
n(s,p)= s^{- \tau} f( (p-p_C) s^{\sigma} )
\end{equation}
therefore in the critical point a power law is expected:
\begin{equation}
n(s,p)= s^{- \tau} f(0)
\end{equation}

Starting from these three last relations, the following relations can be obtained among
 $\sigma$, $\tau$ and the exponents $\beta$ and $\gamma$ :
\begin{equation}
\tau= 2+ \frac{\beta}{\beta + \gamma}
\end{equation}

\begin{equation}
\sigma= \frac{1}{\beta + \gamma}
\end{equation}
In 3D, the best estimation gives $\tau = 2.18$ and $\sigma= 0.45$ \cite{Coniglio}

  Harreis and Bauer introduced a method to deal with
N components percolation, finding new first order phase transitions and 
new empirical relations for the percolation threshold as a function of component
 concentration\cite{Harreis}. Bauer introduced percolation in the study of
 fragmentation ~\cite{bauer_2,bauer3}, for greater details of this computation,
 readers are referred to the article of Stauffer~\cite{stauffer}. We can say that 
a power law arises in Fisher liquid droplet model as well as in percolation.
Therefore we can pose the question whether both belong to the same universality class.
 In both cases, we expect to find finite size and geometry effects. Though we are
 dealing with systems whose physical nature is different, the fact that correlations
 are increased close to the critical point, allows us to consider the possibility of
 both systems belonging to the same universality class.

   L\"ubeck has computed the critical exponents of the order parameter and the
 fluctuations of this parameter, in the case of a conservative lattice, finding
 that the maximum critical dimension for this gas is 4 \cite{Lubeck}. 

Mari ${\it et. al.}$ have performed corrections on the finite system size, for the
  Binder parameter computation of the 3D binomial spins Ising glass \cite{Mari}. 
Janssen ${\it et. al.}$ have shown that for a vector magnetic system of order N,
 when temperature  is considerably greater than the critical temperature, dynamical
 scaling arises in the early evolution as system is suddenly compressed
 up to the critical state \cite{Janssen}.          

   Ying ${\it et. al.}$ have found a nexus between the binding randomness and the 
critical universality in the Potts random binding ferromagnet with a compressed
 disorders trinary distribution in triangular lattices \cite{Ying}.

   Sim\~oes ${\it et. al.}$ have studied the early dynamical evolution of 
the bidimensional Ising model with three spin interactions in a direction, 
considering both Hamiltonian symmetry and boundary conditions when computing
 magnetization, obtaining critical exponents equal to those of four states Potts 
model \cite{Simoes}.

Ying ${\it et. al.}$ have studied the early dynamics and critical universality of
 Potts model with q=2 and q=3 in bidimensional triangular lattices, obtaining the same
 critical exponents as those of a square bidimensional lattice, therefore belonging
 to the same universality class \cite{Ying1}. 

  In this study 3D configurations are randomly generated assigning 
spin values $S=-1,0,1$, with the Hamiltonian:

\begin{equation}
H= -J \sum_{i,j} S_i S_j - H \sum_i S_i .
\end{equation}
 
Spin values are randomly changed in a site and the subsequent change of 
the Hamiltonian is evaluated, maintaining this change with a probability 
equal to:
\begin{equation}
p=\frac{ e^{- \Delta H}/T}{1+ e^{- \Delta H/T}}
\end{equation}
The value of $\tau$ for the colossal magnetoresistance obtained this 
way is equal to $2.39$, cf. fig. 2. Also, we computed an exponent value $\beta= 0.38$, 
therefore colossal magnetoresistance belongs to the same universality class
 as Heavy Ions collisions, as shown in Table 1. The $\beta$ value is close to
 those previously reported by Kudzia ${\it et. al.}$  in Au emulsion fragmentation
 experiments \cite{Kudzia}. Critical temperature for Heavy Ion collisions 
is in the range of 4.5 MeV.

\section{Simulated collisions critical exponent $\tau$ }\label{mol}

  Among the signatures used to study nucleation in nuclear systems, 
 the presence of a maximum in specific heat has been used as a signature 
of a phase transition for periodical systems \cite{Dorso5}. 
Another signature explored is the controversial power law of the 
fragment size distribution, expected to appear close to the 
critical point of a liquid-gas phase transition \cite{Mast1}.
In the relativistic energy range of Heavy Ion collisions, a phase 
transition has been identified with the onset of shock front instabilities
 \cite{Bar1} .

In Heavy Ion collisions, phase diagram is expected to show a first-order 
phase transition turning into a second order phase transition close to the 
critical point. Experimental measurements of Au+Au collision performed
 by GSI Collaboration, lead to a caloric curve giving evidence of 
phase coexistence, in agreement with predictions of statistical 
multifragmentation models excluding volume \cite{Gross1}. 

Other experiments performed in Bevalac extracted  critical exponents and studied
 the dependence between the charge distribution second moment and the biggest fragment
 size as a function of the charged particle multiplicity. Showing that these data are
 consistent with a second order phase transition predicted by the percolation model
 \cite{Bauer1}. 

   L. Shi et. al.  have proved that due to the dependence between asymmetry energy
 and density, some exceeding neutrons belonging to a high density phase must be
 released to the low density phase. The fragment build up in this gas phase, tends
 to  overcome this tendency making the gas phase more similar to the liquid phase and
 reducing the asymmetry in the gas phase. 
An interesting aspect related to phase transition is the mid-rapidity 
region in heavy ion collisions at intermediate energy. In simulated 
semi-peripherical collision simulations, a neck forming region is 
observed that contributes to mean velocity. The low density region in
contact with the high density region (projectile and target) opens 
the possibility for a coexistence between liquid and gas as well as phase conversion
\cite{Shi}.

Since gases do not show first neighbor correlations,
and liquids exhibit a strong two body first neighbor correlation, close to the 
critical point the correlations grow up to include all system particles. 
The behavior close to the critical point is characterized by a loss of the time 
and space scales, what makes that all of the critical phenomena have the same
 characteristics independently of system specifics.

Nevertheless, considering finite and transient systems, the most important scales
 are given by system size and reaction duration though critical phenomena not always
 exist. Hence it is important to prove the possibility of such phenomena in transient
 dynamical systems as those of nuclear reactions.

Concretely, models shall be used to produce similar data resembling a phase
 transition. These data will be dissected to extract, in the best possible way,
 those data produced close to the critical point. The resulting subsets will
 be used to extract the critical exponents. A computation of $\tau$ close to
 the expected values should be an indication of the plausibility of criticality
 in small and transient processes, such as those of heavy ion reactions

\subsection{Molecular Dynamics}

Heavy Ion collisions have been studied with  Molecular Dynamics, 
a method based on Newtonian Mechanics, which is the only one able to describe 
 without adjustable parameters phase transitions, hydrodynamic flow and far from
 equilibrium dynamics. The virtues of Molecular Dynamics to study 
nuclear collisions have been listed in the existing literature~\cite{cherno99}.

In this section, the evolution of two nuclei collision shall be modeled 
with $MD$ (Fig. 5). Nucleons will be treated as point particles subject to potential forces
(Pandharipande Potential). Colliding nuclei are build out as self-assembling particle
 clusters with an spherical geometry. Collision is simulated boosting one of these
 nuclei against the other, and integrating the coupled equations of motion using a
 Verlet algorithm.

Integration precision is ensured demanding energy conservation to a high level. 
Following the dynamics of thousands of collisions for 
different energies, enough information is obtained in order to 
understand ``nuclei'' fragmentation and to obtain a critical exponent 
 $\tau$ for the mass distribution of produced fragments. Nevertheless, to reach 
this stage,  Molecular Dynamics is given in terms of point particles and must be
 transformed in information about fragments. 

\subsection{Fragment Recognition}\label{FR}

In order to convert the particle information provided by Molecular Dynamics in terms
 of fragment information, an Early Cluster Recognition Algorithm is needed, such as
 the one that finds the ``Most bound partition'' of the system~\cite{dorso93}, {\it i.e.}
 the set of clusters $\{ C_i \}$ whose fragment internal energies sum reaches a
 minimum:
\begin{eqnarray}
{ \{C_i\} }   & { = \atop {}} &  { \hbox{argmin} \atop {\scriptstyle \{C_i\}}  } { \textstyle {[E_{ \{C_i\}} = \sum_i E_{int}^{C_i}]} \atop {} } \nonumber \\
E_{int}^{C_i}& = & \sum_i[\sum_{j \in C_i} K_j^{cm} + \sum_{ {j,k
\in C_i}  j \le k} V_{j,k}] \label{eq:eECRA}
\end{eqnarray}
where the first sum is over the partition clusters, 
$K_j^{cm}$ is the kinetic energy of particle $j$ measured in the cluster center of mass
 containing particle $j$, and $V_{ij}$ is the internucleonic potential.
The algorithm used ``simulated annealing'' to find the most bound partition and is known
 as ``Early Cluster Recognition Algorithm'' ($ECRA$). It has been extensively used in
 several fragmentation studies~\cite{cherno99,dorso93,Strachan}, helping to discover
 that excited droplets breakup in a very early stage of collision evolution.

\subsection{Critical exponent $\tau$ for ``realistic'' nuclei}\label{pandha}

A much more realistic model with $MD$, called ``Latino Model'' to 
reflect its Latinamerican origin, was also used. This tri-dimensional model uses
 binary potentials to reproduce the empirical energy and density of nuclear matter,
 as well as realistic effective scattering cross sections.

Nuclei used are spherical droplets with the desired number of protons and 
neutrons, and evolveded to their ``ground state'' using Molecular Dynamics. Once 
the spherical nuclear system is produced at a relatively high temperature, it is 
cooled until it reaches a self-containing state. At this moment, the confining 
potential is removed and the system is cooled until it reaches a reasonable binding energy.

Considering $Ni+Ni$ collisions, projectile is boosted with the desired energy, once 
both projectile and target  have been randomly rotated.
System evolution is reproduced with a Verlet algorithm, ensuring an energy conservation
 better than a $0.01\%$. The required fragment size distribution to compute $\tau$ is 
obtained from information provided by 
$MD$ using $ECRA$, as described in equation~(\ref{eq:eECRA}).
Fig. 6 shows mass spectra obtained from collisions with several energies. 
As can be seen, not all the shapes of these distributions correspond to 
a power law, {\it i.e.} not al the break ups happened in the critical point.
In order to extract the best critical exponent, only those masses in the 
range of $A=2$ to $20$ nucleons were used, {\it i.e.} 
most of the evaporation products and the projectile and target residuals were 
discarded.

Starting from these data, an optimal fit was obtained for 
each projectile energy, with $\tau$ values in the range $2 \sim 3$, 
and for a projectile energy of 1300 MeV, the value obtained is $ \tau = 2.18 $.
This value coincide with the optimal critical exponent for the 
percolation model, as mentioned in section 3. The observed values of $\tau$ provide
 enough evidence to consider critical nucleation and confirm other studies of
 Ni+Ni central collision, where the critical behavior was detected in this projectile
 energy range \cite{oaxt2001}.

\section{Conclusions}\label{CP}

 Table 1 shows all values of $\tau$ obtained for these computations 
and compared with others reported elsewhere and obtained via percolation 
as well as with others simulated via bidimensional Molecular Dynamics
 with a Lennard-Jones potential \cite{BarraHI}. Results from both $MD$ collisions 
and genetic algorithm replicating colossal magnetoresistance, show that 
critical nucleation occurs in transient and static small fragmenting systems,
 and that these systems belong to the same universality class.

\begin{eqnarray*}
\text{\bf TABLE 1 }  \\ \hline
\text{\bf Nucleation Model}& \qquad \text{\bf $\tau$ Value }  \\ \hline
\text {Colossal Magnetoresistance} & \qquad \text{ $2.38 $ }\\  \hline
\text{3-D MD Collision Simulations } & \qquad \text{$2.18$ }\\ \hline
\text{Cubic Lattice Percolation} \cite{BarraHI} 
 & \qquad \text{$2.32 \pm 0.02$ }\\ \hline
\text{Spherical Lattice Percolation \cite{BarraHI}} & \qquad \text{$2.20 \pm 0.1$ }\\ \hline
\text{2-D MD Collision Simulations \cite{BarraHI}} & \qquad \text{$2.32 \pm 0.02$ }\\ \hline
\end{eqnarray*}

\section{Acknowledgements}

 Work financed by National Science Foundation (PHY-96-00038),
Universidad de Buenos Aires (EX-070, Grant No. TW98,
CONICET Grant No. PIP 4436/96) and Universidad Aut\'onoma
Metropolitana-Azcapotzalco (Supercomputing Lab).

\end{document}